# CROSS LAYER AWARE ADAPTIVE MAC BASED ON KNOWLEDGE BASED REASONING FOR COGNITIVE RADIO COMPUTER NETWORKS

Vibhar Pathak[1,] Dr. Krishna Chandra Roy[2] and Santosh Kumar Singh[3]

[1]Department of Information Technology, Suresh Gyan Vihar University, Jaipur, India
vibhakar_p@yahoo.com
[2]Department of ECE, SBCET, Benad Raod, Jaipur, India
roy.krishna@rediffmail.com
[1]Department of Computer Engineering, Suresh Gyan Vihar University, Jaipur, India
sksmtech@yahoo.com

*ABSTRACT :*

*In this paper we are proposing a new concept in MAC layer protocol design for Cognitive radio by combining information held by physical layer and MAC layer with analytical engine based on knowledge based reasoning approach. In the proposed system a cross layer information regarding signal to interference and noise ratio (SINR) and received power are analyzed with help of knowledge based reasoning system to determine minimum power to transmit and size of contention window, to minimize backoff, collision, save power and drop packets. The performance analysis of the proposed protocol indicates improvement in power saving, lowering backoff and significant decrease in number of drop packets. The simulation environment was implement using OMNET++ discrete simulation tool with Mobilty framework and MiXiM simulation library.*

*KEYWORDS :*

*Cognitive radio, SDR, SINR, Received Power, MAC, OMNET++*

## I. INTRODUCTION:

There is a significant amount of unused space (or white space) in the licensed radio spectrum due to non-uniform spectral demand in time, frequency and space and static spectrum allocation policies widely used today. Studies sponsored by FCC in [12] show that over 70% of the allocated spectrum is not in use at any time even in a crowded area where the spectral usage is intensive. On the other hand, the remaining portion of the unlicensed spectrum (e.g. the ISM band) is being exhausted by emerging wireless services and applications, leading to the so-called spectral scarcity problem. One solution to this problem is to allow unlicensed spectrum users to use the white space and keep the interference to licensed users below an acceptable level. This is called the dynamic spectrum access (DSA) scheme, which can be realized by cognitive radio (CR) techniques [13]–[16] A CR device monitors a swath of spectrum including those occupied by licensed services and attempts to identify the "white" space (or the spectrum hole), which is referred to as the idle period between consecutive accesses of licensed users, and exploits it for communication at that specific geographical location.

Recent development in slicon technology leads to development of smart reprogrammable circuits Using which a new class of intelligent or "COGNITIVE" radios can be develop based on Software Defined Radio(SDR).Such radio based system would be capable of dynamic physical adaptation. In recent past development of cognitive radio hardware and software, especially at the physical layer has received considerable attention., the question how one can transform a set of cognitive radio into a cognitive network is less considered by research community. Cognitive radio or agile radio is a technology to choose a wide variety of radio parameters and protocol standard in adaptive manner on observed radio link and network conditions. Such type of system is capable of handling cross layer parameter change and advice the network to change the system with view of minimum power consumption, lowering back off and reducing rate of drop packets, and hence updating utilization of network resource. In generalized case the cognitive radio is capable of adapting modulation of waveform, OSA (opportunistically spectrum access), MAC protocols, network protocols. The cognitive radio can make runtime change to protocols to avoid collisions by transmitting packets with minimum power utilized for hop-to-hop transfer. There have been many research works addressing physical layer agility of cognitive radio system based on OSA [2][3]. This paper proposes novel Cross Layer aware adaptive MAC protocol based on knowledge–based reasoning for cognitive radio. In this different type of Mac protocol can be adapted by the system on basis different physical parameter, receiver power, SINR and minimum power to transmit for hop-by-hop packet transfer knowledge base. The candidate MAC protocols have different advantages in various network situations. The proposed CLA-AMAC will provide a framework to observe and make decision to switch on between those candidate protocols. The switching between different MAC protocols will be validated and used in future by the system based on short-term statistics.

The proposed system (CLA-AMAC) is based on the "COGNET" cognitive radio protocol architecture described in[4].The COGNET system includes the concept of "Global Control Plane" GCP which supports exchange of control information between networking cognitive radio devices. The MAC adaptation requires cross-layer parameters and tokens to maintain protocol consistency between radio nodes sharing the channels. The GCP is also used in many cross-layer protocols including routing in MANET.

In this paper we will focus on the simulation of CLA-AMAC using open source network simulation tools OMNET + + with mobility framework. The candidate protocols used in our protocols are commonly used protocol in wireless networks; these are CSMA/CA, TDMA and multi-channel Wi-Fi protocol. CSMA is ideal for short burst source with light to medium traffic volume. In streaming source TDMA protocol do well and Multi-channel Wi-Fi protocol is ideal for real-time mission critical application. We evaluate the proposed CLA-AMAC protocols with various traffic types.

In next section we begin by describing related work followed by the protocol is discussed in section III finally we provide Simulation results in section IV and future work in section V.

## II. Related Works.

The benefit of separate control and data planes in COGNET in the COGNET architecture have been previously explained in[4]. The GCP uses low-rate radio PHY with wide coverage for robustness and may include protocol modules for topology discovery, bootstrapping, address assignment. The data plane protocol stack supports data communication via PHY MAC and routing modules specified through an API that interacts with control modules in the radio node. The GCP control architecture is used for our CLA-AMAC.

Many approaches have already been proposed to reduce the number of collisions by substituting the binary exponential back-off algorithm of the IEEE 802.11 by novel back-off approaches or selecting an intermediate value instead of resetting the CW value to its initial value or some random value of CW upper or lower bound. The most related work to our back-off mechanism is the determinist contention window algorithm (DCWA) in [17]. DCWA increases the upper and lower bounds instead of just doubling the CW value. In each contention stage, a station draws a back-off interval from a distinct back-off range that does not overlap with the other back-off ranges associated to the other contention stages. In addition, the back-off range is readjusted upon each successful transmission by taking into account the current network load and history (resetting the back-off ranges mechanism; see details in [17]). Among the related work concerning energy conservation, such as power saving or power control mechanisms, the power saving mechanism (PSM) is the most familiar. It is provided by the standard [1], which allows a node to go into doze mode. Power control schemes, varying the transmit power in order to reduce the energy consumption, have already been presented in many studies; for example, see [18–22]. These schemes and many others have shown that power control protocols can achieve a better power conservation and higher system throughput through a better spatial reuse of the spectrum.

One part of the research in the field of power control focuses on dependencies and tradeoffs between both the transmit power and the carrier sense threshold, while another part focuses only on the adjustment of the carrier sense threshold [23] The work in [12] investigated the tuning of the transmit power, carrier sense threshold, and data rate in order to improve spatial reuse. The authors have shown that tuning the transmit power is more advantageous than tuning the carrier sense threshold. Cross-layer protocols contributing to the enhancement of the MAC layer and the adjustment of the power level have also been presented in many papers. One of them, the power adaptation for starvation avoidance (PASA) algorithm [17], was designed following the observation from that the request-to-send/clear-to-send (RTS/CTS) collision avoidance mechanism of the IEEE 802.11 DCF cannot eliminate collisions completely. This can lead to a channel capture where a channel is monopolized by a single or a few nodes. The authors of [17] studied how to control the transmission power properly in order to offer a better fairness and throughput by avoiding a channel capture. The power level increases exponentially and decreases linearly in the PASA, while using an RTS/CTS control scheme. PASA is not applicable with the basic access scheme. It requires that a neighbor power table (NPT) is maintained by each node with information such as the minimum power that must be maintained according to the distance to the destinations, which should be obtained through some location service. After all, maintaining the NPT table with "fresh" data is not realistic in a mobile ad hoc environment taking into account interferences, fading effects, movement of the nodes, and deaths and new entries of nodes. The carrier sense multiple access protocol with power back off (CSMA/PB) has been presented in [18]. The CSMA/PB reduces the transmission power level in order to avoid collisions, following the observation that, in a smaller transmission area, interferences and contentions are expected to be reduced. Results obtained in [18] are based on an optimistic centralized power-aware routing strategy, which illustrates the potential of the power back off. The CSMA/PB protocol has been evaluated with three transmission power levels only, thus the amount of power decreases fast. Therefore, it is really important that the routing protocol takes power levels into account. Each node has to maintain the routing table with entries for each destination with corresponding power levels.

## III CLA-AMAC PROTOCOL

The goal of the CLA-AMAC protocol is to save energy (which leads to an extension of the lifetime of nodes) and to reduce the number of collisions. However, the CLA-AMAC protocol does not degrade the throughput performance in terms of the throughput and data rate, while fulfilling these goals. The CLA-AMAC protocol tackles a couple of problems that exist in the current implementation of the standard. It does this by two means; first it concentrates on the flexible adjustment of the upper and lower bounds of the CW to lower the number of collisions. Secondly, it uses a power control scheme to limit the waste of energy and also to lower the number of collisions. Hence, it has MAC-PHY cross layer architecture. To tackle the inefficient use of the back-off window in the standard, we developed a MAC protocol that makes use modified enhanced selection Bounds algorithm (EsB) [19]). The mEsB (modified Enhanced selection Bounds) adjusts the lower and upper bounds of the CW range, taking into account the number of retransmissions attempts, Received power from neighbor, and SINR sensed. Each node can predict number of active 1- hop neighbor, based on successfully detected signals Knowledge based reasoning engine. In [20] the utilization rate of the slots (slot utilization) observed on the channel by each station is used for a simple, effective and low-cost load estimate of the channel congestion level. The protocol uses knowledge based reasoning system generate collision history and predict the same During the resetting stage, the CW value is reset to a value which depends on the predicted collisions. This forms the MAC part of the CLA-AMAC protocol and results in a reduction of the number of collisions.

The goal is not only to lower the number of collisions, but also to save energy. If we reflect on the reason why messages collide, it becomes clear that this is because too many nodes are too close to each other. They could be positioned a few meters from each other, but their transmission range is far greater than these few meters. Hence, the nodes are too close to each other relative to their respective transmission range.
This not only results in a higher number of collisions, but also in an excessive use of energy to transmit a packet.

However, not receiving an acknowledgment for a sent packet does not always mean that the packet was lost or corrupted because there was too much interference. It could also happen that the transmission power was simply too low to reach any of the surrounding nodes. Therefore, the CLA-AMAC protocol takes the signal-to-interference-and noise ratio (SINR) into account. If no acknowledgment has been received, but the noise level (deducted from the SINR) is low, then we assume that the transmission power was too low to reach any of the neighbors. In that case the transmission power is increased.

The CLA-AMAC power control part is based on this observation of received power of packet from 1-hop neighbor and it lowers its transmission power (while observing too high noise in the vicinity) when it does not get the acknowledgment that a packet has been received successfully. The final result will be that all nodes will find their optimal transmission power that ensures that they can reach their neighbors, but not interfere with other nodes. The whole information regarding SINR, minimum power to transmit, receiver power will be communicated to whole network via GCP and reasoning will be developed using knowledge-based system to predict the behavior of network.

## IV. SIMULATION RESULTS

The proposed cross-layer protocol has been implemented in the OMNET++ 4.0 network simulator [10]. The simulations have been carried out for various topologies, scenarios with different kinds of traffic, and routing protocols. The following performance metrics have been used:
(i) Total packets received,
(ii) Average throughput (Mbps),
(iii) Lifetime LND (seconds),
(iv) FND: first active node died (seconds),
(v) Lifetime RCVD (seconds),
(vi) Average aggregate delay (seconds),
The first node died metric is defined as the instant in time when the active (a node transmitting/receiving) first node died. We have defined the network lifetime as the time duration from the beginning of the simulation until the instant when the active (a node transmitting/receiving) last node died, that is, there is no live transmitter-receiver pair left in the network. The Lifetime RCVD is specified as the instant in time when the last packet is received.

The average throughput has been defined as Threshold = Total number Packets received Simulation Time [Mbps] and average sending bit rate has been defined as bit = Total number Packets sent Simulation Time [Mbps].

Table 1

| Parameter | Value |
| --- | --- |
| Number of active nodes | 25, 50 *(default)* |
| Simulations area | ≤ 1000 * 1000m |
| Topology | Random |
| PHY/MAC | DSSS, IEEE 802.11b |
| SINR thr. (dB) | 22.05 |
| Type of netwok | *homo/hetero*-geneous |
| Initial energy (J) variable = | 0.5–..., 5, 20 |
| $Pt_{MAX}$ – | 250m 0.200888W |
| $Pt_{MAX}$ – | 100m 0.010072W |
| *txPower$_{init}$* | 250 _ 100 meters |
| rxPower | 45% of $Pt_{MAX}$ |
| idlePower | 30% of $Pt_{MAX}$ |
| Capture Thr.(dB) | 10 |
| Traffic model | CBR/UDP |
| Payload size (bytes) | 2048 _ 100–8192 |
| $CW_{min}$ – $CW_{max}$ (slots) | 15–1023 |
| Simulation time (s) | ≤650 |
| Routing | AODV *(default)*, DSR, OLSR |
| Movement | Random and constant |
| Mobility model | Turtle Model |
| Speed (m/s) | 0 – 2_ ≤20; 1.5– *(default)* |
| Access scheme Basic | *(default)* _ RTS/CTS |

Table 2: Typical values of path loss exponent and shadowing deviation.

| Environment | $\rho$ (dB) | $\sigma$ (dB) |
| --- | --- | --- |
| Outdoor Free space | 2.4 | 4 to 12 |
| Outdoor Shadowed Urban | 2.7 to 5.6 | 4 to 12 |
| Indoor Line-of-sight | 1.6 to 1.8 | 3 to 6 |
| Indoor Obstructed | 4 to 6 | 6.8 |

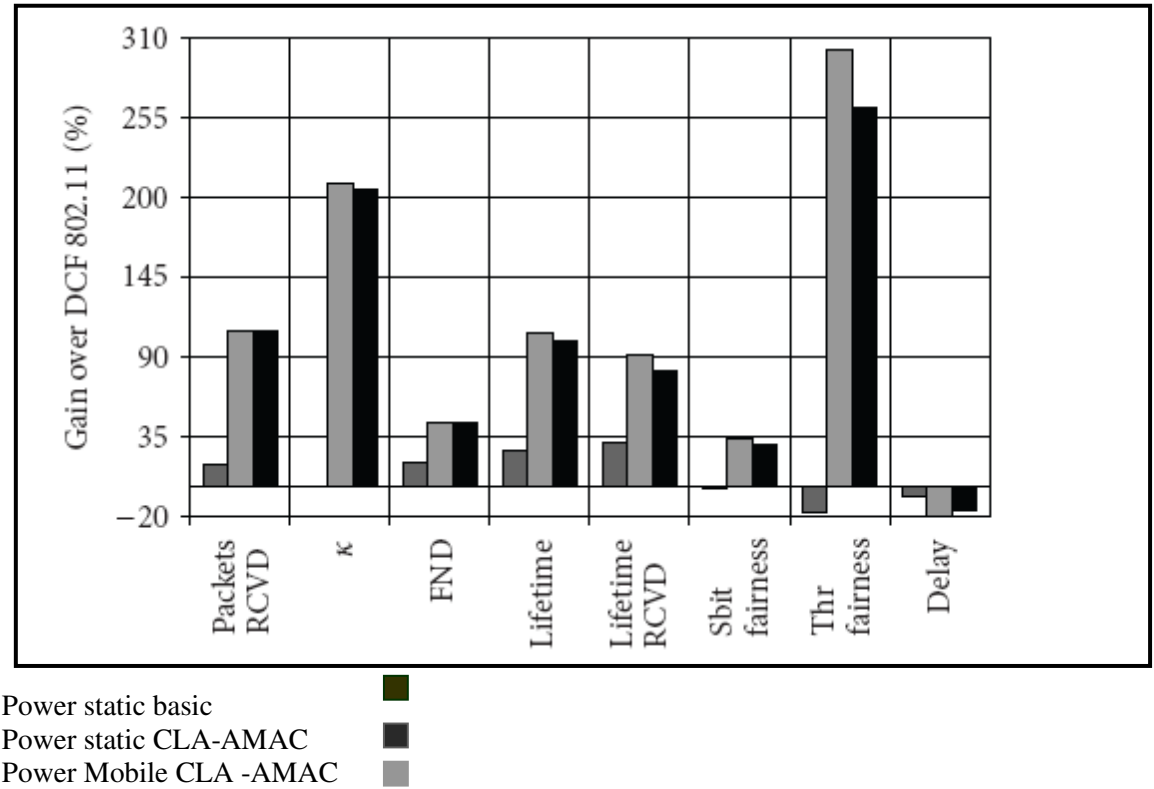

Power static basic
Power static CLA-AMAC
Power Mobile CLA -AMAC

Figure 1

First, we defined a simulation scenario with 40 static nodes randomly distributed in a shadowed urban area where nodes send a CBR packet (2048 bytes payload size) from the beginning till the end of the simulation every 0.025 seconds. Figure 4 depicts the number of collisions per node in one of the simulation scenario runs (20 simulation runs in total). Notice that with the CLA-AMAC protocol most of the nodes have much fewer collisions, although the lifetime of the network is increased significantly (See Figure 6). Figure 5 shows the total number of packets received by the DCF standard, *basic* power control protocol, and CLA-AMAC protocol. The tuning of the CLA-AMAC protocol has been investigated as can be observed in the figure. The *CLA-AMAC,* the protocol outperforms the IEEE 802.11 DCF standard and *basic* power control protocol noticeably. The *CLA-AMAC achieves* the best performance, which means that the history of collisions experienced has an influence in a static environment.

Figure1 shows the gain in percentage over the IEEE 802.11 DCF standard obtained by the *basic* power control protocol in the static network and the CLA-AMAC protocol in both static and mobile networks. Note that, thanks to PHY (power level adjustment) and MAC (recovery mechanism and CW resetting) layer treatment, the number of collisions can be decreased noticeably while saving lot of the energy which leads to an increase of the lifetimes (LND and lifetime RCVD) of the network and the throughput. The performance of the Lifetime RCVD is worse than the performance of the lifetime of the network, which means that some last transmitter-receiver pairs still have connections; however, the packets cannot be routed to the destination. The performance of the throughput fairness, which is improved tremendously, is explainable since nodes give others more opportunity to access a wireless channel while decreasing the transmit power level. On the other hand, by increasing the power (upon a consecutive collision and too low noise in the vicinity), their chance to get to the channel is increased since their coverage transmit area is wider. However, the average delay is degraded, because the CLA-AMAC protocol adjusts both the lower and upper bounds of the CW range and allows to decrease (apart from an increase) the power level, which in consequence can increase the average delay. In this work we have designed a novel cross-layer protocol, Adaptive protocol. The protocol adjusts the upper and lower bounds of the contention window to lower the number of collisions. Secondly, it uses a power control scheme, triggered by the MAC layer, to limit the waste of energy and also to decrease the number of collisions. Apart from that the protocol uses knowledge based reasoning for prediction SINR and receiver power for further optimization of contention window and further decrease in number of collision   The protocol has been evaluated in three different scenarios and compared to the IEEE 802.11 DCF standard and the basic power control protocol.

**Vibhakar Pathak** received his Master of Computer Application from Indira Gandhi Open University, India. He having 12 year teaching experience and pursuing his Ph.D. degree in Engg. at the School of Engineering, Suresh Gyan Vihar University, Jaipur, India. He is also presented several paper in International and National conference. His current research interests include next generation mobile wireless communication.

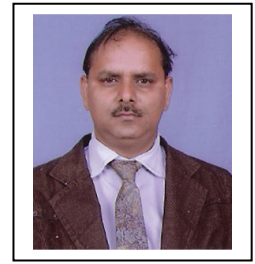

**Dr. Krishna Chandra Roy** received his M.Sc.(Engg) degree in Electronics and Communication Engineering from NIT Patna, Bihar, India and Ph.D degree in "Digital Signal Processing in a New Binary System" year 2003. He has currently Reader & Head of department (P.G.Engg.) of ECE, Sri Balaji College of Engg. & Tech., Jaipur, India and having 13 year teaching experience. He published two books Problems & solution in Electromagnetic Field Theory by Neelkanth Publishers (p) Ltd., Year-2006 and Digital Communication by University Science Press, Year-2009 respectively. He is also published and presented several paper in International and National conference. His current research interests include Digital Signal Processing and embedded system.

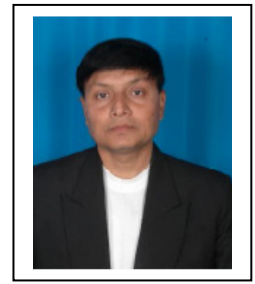

**Santosh Kumar Singh** received his B.E. degree in Electronics and Communication Engineering from S. J. College of Engineering, under Mysore University, Karnataka, India, year 1995 and M.Tech in Information Technology in 2004. He having 13 year teaching experience and pursuing his Ph.D. degree in Engg. at the School of Engineering, Suresh Gyan Vihar University, Jaipur, India. He published one book and one paper in well-reputed publication. He is also presented several paper in International and National conference. His current research interests include next generation wireless networks, wireless sensor networks and industrial embedded system.

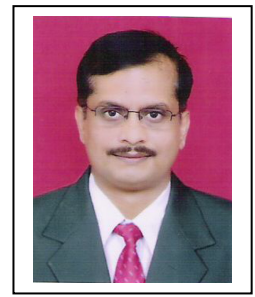